\documentclass[aps]{revtex4}
\usepackage{eurosym}
\usepackage{amsfonts}
\usepackage{amsmath}
\usepackage{amssymb,epsf}
\usepackage{color}
\usepackage{graphicx}
\usepackage{epstopdf}
\usepackage{float}
\usepackage{caption}
\usepackage{subfig}

\begin{document}

\title{Entropy spectrum of charged BTZ black holes in massive gravity's
rainbow}
\author{B. Eslam Panah$^{1,2}$\footnote{%
email address: beslampanah@shirazu.ac.ir}, S. Panahiyan$^{3}$ \footnote{%
email address: sh.panahiyan@gmail.com} and S. H. Hendi $^{1,2}$ \footnote{%
email address: hendi@shirazu.ac.ir} }
\affiliation{$^1$ Physics Department and Biruni Observatory, College of Sciences, Shiraz
University, Shiraz 71454, Iran\\
$^2$ Research Institute for Astronomy and Astrophysics of Maragha (RIAAM),
P.O. Box 55134-441, Maragha, Iran\\
$^3$ Helmholtz-Institut Jena, Fr\"{o}belstieg 3, Jena D-07743 Germany}

\begin{abstract}
Regarding the significant interests in massive gravity and combining it with
gravity's rainbow and also BTZ black holes, we apply the formalism
introduced by Jiang and Han in order to investigate the quantization of the
entropy of black holes. We show that the entropy of BTZ black holes in
massive gravity's rainbow is quantized with equally spaced spectra and it
depends on the black holes' properties including massive parameters,
electrical charge, the cosmological constant and also rainbow functions. In
addition, we show that quantization of the entropy results into the
appearance of novel properties for this quantity such as; the existence of
divergencies, non-zero entropy in vanishing horizon radius and possibility
of tracing out the effects of black holes' properties. Such properties are
absent in the non-quantized version of these black holes' entropy.
Furthermore, we investigate the effects of quantization on the
thermodynamical behavior of the solutions. We confirm that due to
quantization, novel phase transitions points are introduced and stable
solutions are limited to only dS black holes (AdS and asymptotically flat
solutions are unstable).
\end{abstract}

\maketitle

\section{Introduction}

General relativity (GR) is a successful theory of gravity with certain
shortcomings. For example: accelerated expansion of the universe, massive
gravitons and the\ ultraviolet (UV) behavior could not be explained with GR.
To address these issues and other ones, GR should be modified. There are
some modified theories such as; Horava-Lifshitz gravity \cite%
{Horava1,Horava2}, gravity's rainbow \cite%
{Rainbow1,Rainbow2,Rainbow3,Rainbow4} and also massive gravity \cite%
{Mass1,Mass2,Mass3,Mass4,Mass7,Mass8,Mass9}.

In order to understand the UV behavior of GR, various attempts have been
made to obtain different models of UV completion of GR such that\textbf{\ }%
they should reduce to GR in the infrared (IR) limit. The first attempt in
this field is related to Horava-Lifshitz gravity \cite{Horava1,Horava2}, in
which space and time are made to have different Lifshitz scaling. Although
this theory reduces to GR in the IR limit, its behavior is different from
that of GR in the UV regime. It is notable that, Horava-Lifshitz gravity is
based on a deformation of the usual energy-momentum dispersion relation in
the UV limit, in which it reduces to the usual energy-momentum dispersion
relation in the IR limit. Another approach for extracting the UV completion
of GR is called gravity's rainbow \cite{Rainbow1}. The theory is based on
the deformation of the usual energy-momentum dispersion relation in the UV
limit, and similar to Horava-Lifshitz gravity, gravity's rainbow reduces to
GR in the IR limit.\ It was shown that the quantum corrections in a
gravitational system could be observed in dependency of its space-time on
the energy of particles probing it which is gravity's rainbow point of view
\cite{Smolin,Garattini}. Also, by considering a suitable choice of the
rainbow functions, the Horava-Lifshitz gravity can be related to gravity's
rainbow \cite{GarattiniS}. This is because both of these theories are based
on modifying the usual energy-momentum dispersion relation in the UV limit.
It is worthwhile that such a modification of the usual energy-momentum has
also been obtained in discrete spacetime \cite{Hooft}, the spin-network in
loop quantum gravity (LQG) \cite{Gambini}, spacetime foam \cite%
{Amelino-Camelia}, ghost condensation \cite{Faizal}, and non-commutative
geometry \cite{Carroll}. The non-commutative geometry occurs due to
background fluxes in string theory \cite{Seiberg,Cheung}, and it is used to
derive one of the most important rainbow functions in gravity's rainbow \cite%
{Amelinof(e),Jacob}.

\textbf{\ }In other words, the geometry of spacetime is modified to be
energy dependent and this energy dependency of the metric is incorporated
through the introduction of rainbow functions. The standard energy-momentum
relation in gravity's rainbow is given as
\begin{equation}
E^{2}f^{2}(E/E_{P})-p^{2}g^{2}(E/E_{P})=m^{2},  \label{MDR}
\end{equation}%
in which $E$ and $E_{P}$ are the energy of test particle and the Planck
energy, respectively. For the sake of simplicity, we will use $\varepsilon
=E/E_{P}$. Also, $f(\varepsilon )$ and $g(\varepsilon )$ are energy
functions which are restricted as $\lim\limits_{\varepsilon \rightarrow
0}f(\varepsilon )=1$ and $\lim\limits_{\varepsilon \rightarrow
0}g(\varepsilon )=1$, in the IR limit and could be used to build an energy
dependent metric with the following recipe

\begin{equation}
\hat{g}(\varepsilon )=\eta ^{ab}e_{a}(\varepsilon )\otimes e_{b}(\varepsilon
),  \label{rainmetric}
\end{equation}%
where
\begin{equation}
e_{0}(\varepsilon )=\frac{1}{f(\varepsilon )}\tilde{e}_{0},\qquad
e_{i}(\varepsilon )=\frac{1}{g(\varepsilon )}\tilde{e}_{i},
\end{equation}%
in which $\tilde{e}_{0}$ and $\tilde{e}_{i}$ are related to the energy
independent frame fields. It is notable that $E$ can not exceed $E_{P}$, so,
$0<\varepsilon \leq 1$. In other words, the gravity's rainbow produces a
correction to the metric that becomes significant when the particle's energy
approaches the Planck energy.\ It is notable that, there are three models
for these energy functions as:

Case I) it is related to the hard spectra from gamma-ray bursts \cite%
{Amelinof(e)}, with the following form%
\begin{equation}
f\left( \varepsilon \right) =\frac{e^{\beta \varepsilon }-1}{\beta
\varepsilon },~~~~\&~~~~g\left( \varepsilon \right) =1.
\end{equation}

Case II)\ it is motivated by studies conducted in loop quantum gravity and
non-commutative geometry \cite{Jacob} as%
\begin{equation}
f\left( \varepsilon \right) =1,~~~~\&~~~~g\left( \varepsilon \right) =\sqrt{%
1-\eta \varepsilon ^{n}}.
\end{equation}

Case III)\ it is due to the consideration of constancy of the velocity of
light \cite{Magueijof(e)}%
\begin{equation}
f\left( \varepsilon \right) =g\left( \varepsilon \right) =\frac{1}{1-\lambda
\varepsilon },
\end{equation}%
in which in the above models $\beta $, $\eta $\ and $\lambda $\ are
constants which could be determined by experiment.

In the gravity's rainbow context and by combining various gravities, the
black hole and cosmological solutions have been studied in some literatures.
For example, $F(R)$ gravity's rainbow \cite{F(R)I}, Gauss-Bonnet gravity's
rainbow \cite{fai}, dilatonic gravity's rainbow \cite{dilatonII} and
Galileon gravity's rainbow \cite{GalileonI} have been investigated. Also, in
ref. \cite{NonsingularI}, it was shown that by considering a special limit
on rainbow functions, we encounter with nonsingular universes in gravity's
rainbow. The remnant of the black objects and also the absence of black
holes at LHC due to gravity's rainbow have been evaluated in refs. \cite%
{remnantI,remnantII,remnantIII}. Modified TOV in gravity's rainbow and
investigating the properties of magnetic neutron stars and dynamical
stability conditions have been perused \cite{TOVI,TOVII}. Heat engine and
geometrothermodynamics of the obtained black holes in gravity's rainbow have
been studied in ref. \cite{heatRainbow}.

On the other hand and in order to have massive gravitons, GR must be
modified, because the gravitons are massless particles in GR. Therefore,
Fierz and Pauli were the first to study the theory describing the massive
gravitons (FP massive theory) \cite{Mass1,Mass2}. Studies done by van Dam,
Veltman and Zakharov showed that FP massive theory encounter with a
discontinuity (vDVZ discontinuity) \cite{vDVZI,vDVZII,vDVZIII}. In order to
remove this problem, one has to generalize the linear FP massive gravity to
a nonlinear one. Later, Boulware and Deser found out that this theory of
massive gravity suffers a ghost instability at the nonlinear level \cite%
{Mass3,Mass4}. Recently, there has been a great interest in the modification
of GR on the nonlinear level to include massive gravitons. Among the studies
done in this regard, de Rham, Gabadadze and Tolley (dGRT) were able to
introduce an interesting massive gravity without any ghost in arbitrary
dimension \cite{dRGTI,dRGTII}, in which a stable nonlinear massive gravity
\cite{Mass7,Mass8,Mass9} was employed to conduct the investigations.
Recently, several interesting black hole solutions have been obtained in
various massive gravities \cite%
{BHdiffmassI,BHdiffmassII,BHdiffmassIII,BHdiffmassV,BHdiffmassVII,BHdiffmassVIII,BHdiffmassIX,BHdiffmassXI,BHdiffmassXIII,BHdiffmassXIV,BHdiffmassXV,BHdiffmassXVI,BHdiffmassXVII}%
. Charged black holes and their thermodynamics in massive gravity have been
evaluated in refs. \cite{ThMassiveI,ThMassiveIV,ThMassiveV,ThMassiveVI}. Van
der Waals like phase transition and geometrical thermodynamics of black
holes in massive gravity have been investigated in refs. \cite%
{PVI,PVII,PVIII,PVIV,PVV,PVVI,PVVIII,PVIX}. Modified TOV in massive gravity
and investigating the properties of neutron stars and also dynamical
stability conditions have been done \cite{TOVMassI,TOVMassII}. Black holes
as heat engine in massive gravity have been investigated in refs. \cite%
{HeatI,HeatII}. Considering massive gravity the white dwarfs have been
studied in ref. \cite{whiteMass}

The first three dimensional black hole solution in the presence of the
cosmological constant was obtained by Ba\~{n}ados, Teitelboim, and Zanelli
which is known as BTZ black hole \cite{btz}. Later, it was shown that these
solutions have central roles in understanding several issues such as black
hole thermodynamics \cite{car1,ast,sar}, quantum gravity, string theory, the
anti-de Sitter spaces/ conformal field theories (AdS/CFT) conjecture \cite%
{wit,car2} and investigation of gravitational interaction in low dimensional
spacetime \cite{wit1}. The charged BTZ black hole is the correspondence
solution of Einstein-Maxwell gravity in three dimensions \cite{car1,mar,cle}%
. Recently, charged BTZ black holes with two generalizations of the massive
gravity and gravity's rainbow have been studied \cite%
{CaiMassive,hend1,BTZMassive}.

As the first cornerstone, the concept of Hawking radiation of black holes
improved our knowledge toward the quantum theory of gravity. Then,
Bekenstein showed that there is a lower bound for the event horizon area of
black holes as \cite{Bekenstein1972}
\begin{equation}
\left( \Delta A\right) _{\min }=8\pi l_{p}^{2},
\end{equation}%
in which $l_{p}$ is the Planck length. It is notable that this lower bound
does not depend on the parameters of black holes. On the other hand,
quasinormal mode (QNM) frequencies are known as the characteristic sound of
black hole. These QNMs should have an adiabatic invariant quantity. Hod
extracted the area and also entropy spectrum of black hole from QNMs \cite%
{HodI,HodII}. Using Bohr-Sommerfield quantization rule ($I_{adiabatic}=n%
\hbar $), Hod showed that the area spectrum of Schwarzschild black hole is
equispaced. Using the well known Bekenstein-Hawking area law and considering
the area spectrum, one can obtain the entropy spectrum of black holes as $%
\Delta S_{bh}=\ln 3$. On the other hand, Kunstatter obtained the area
spectrum of higher dimensional spherical symmetric black holes by
considering the adiabatic invariant quantity in the following form \cite%
{Kunstatter}
\begin{equation}
I_{adiabatic}=\int \frac{dE}{\Delta \omega \left( E\right) },
\end{equation}%
where $\Delta \omega =\omega _{n+1}-\omega _{n}$, $E$ and $\omega $ are the
energy and frequency of QNM, respectively. Later, Hod and Kunstatter
calculated the area spectrum by considering the real part of QNM frequency.
Next, Maggiore \cite{Maggiore} refined Hod's idea by proving that the
physical frequency of QNM is determined by its real and imaginary parts. A
new method was proposed by Majhi and Vagenas in order to quantize the
entropy without using QNM. They used the idea of relating an adiabatic
invariant quantity to the Hamiltonian of the black hole, and then obtained
an equally spaced entropy spectrum with its quantum to be equal to the one
obtained by Bekenstein \cite{Majhi}. In the tunnelling picture, we can
consider horizon of black hole to oscillate periodically when the particle
tunnels in or out of black hole. Therefore, we can use this viewpoint and
consider an adiabatic invariant quantity as (or action of the oscillating
horizon)
\begin{equation}
I=\int p_{i}dq_{i},  \label{ActionI}
\end{equation}%
in which $p_{i}$ is the corresponding conjugate momentum of the coordinate
of $q_{i}$ ($i=0,1$ where $q_{0}=\tau $ and $q_{1}=r_{h}$, in which $\tau $
and $r_{h}$ are related to the Euclidean time and the horizon radius,
respectively). By using the Hamilton's equation ($\overset{.}{q}_{i}=\frac{dH%
}{dp_{i}}$), one can rewrite the equation (\ref{ActionI}) as%
\begin{equation}
I=\int \int_{0}^{H}dHd\tau +\int \int_{0}^{H}\frac{dH}{\overset{.}{r_{h}}}%
dr_{h}=2\int \int_{0}^{H}\frac{dH}{\overset{.}{r_{h}}}dr_{h},
\label{Action2}
\end{equation}%
where $H$ is the Hamiltonian of system and $\overset{.}{r_{h}}=\frac{dr_{h}}{%
d\tau }$. Now, we want to calculate the above adiabatic invariant quantity,
so we consider a static metric in gravity's rainbow as%
\begin{equation}
ds^{2}=-\frac{\psi \left( r,\varepsilon \right) }{f^{2}\left( \varepsilon
\right) }dt^{2}+\frac{1}{g^{2}\left( \varepsilon \right) }\left[ \frac{dr^{2}%
}{\psi \left( r,\varepsilon \right) }+r^{2}d\varphi ^{2}\right] .
\label{metric1}
\end{equation}

It is notable that, we can obtain $r_{h}$ by using $\psi \left(
r_{h},\varepsilon \right) =0$. Finding the oscillating velocity of black
hole horizon, we can calculate the equation (\ref{Action2}). In the
tunnelling picture, when a particle tunnels in or out, horizon of black hole
will expand or shrink due to gaining or losing the mass in the black hole.
Since the tunnelling and oscillation happen simultaneously, the tunnelling
velocity of particle is equal and opposite to the oscillating velocity of
black hole horizon ($\overset{.}{r_{h}}=-\overset{.}{r}$). Also, we have to
Euclideanize the introduced metric (\ref{metric1}) by using the
transformation $t\rightarrow -i\tau $. So, we have
\begin{equation}
ds^{2}=\frac{\psi \left( r,\varepsilon \right) }{f^{2}\left( \varepsilon
\right) }d\tau ^{2}+\frac{1}{g^{2}\left( \varepsilon \right) }\left[ \frac{%
dr^{2}}{\psi \left( r,\varepsilon \right) }+r^{2}d\varphi ^{2}\right] .
\end{equation}

It is notable that, when a photon travels across the horizon of black hole,
the radial null path (or radial null geodesic) is given by
\begin{equation}
ds^{2}=d\varphi ^{2}=0~\rightarrow ~\overset{.}{r}=\pm i\left( \frac{g\left(
\varepsilon \right) \psi \left( r,\varepsilon \right) }{f\left( \varepsilon
\right) }\right) ,  \label{radialnull}
\end{equation}%
in which the negative sign denotes the incoming radial null paths and also
the positive sign represents the outgoing ones. It is notable that, we
consider the outgoing paths (the positive sign of Eq. (\ref{radialnull})) in
order to calculate the area spectrum, because these paths are more related
to the quantum behaviors under consideration. So, the shrinking velocity of
black hole horizon is given by%
\begin{equation}
\overset{.}{r_{h}}=-\overset{.}{r}=-i\left( \frac{g\left( \varepsilon
\right) \psi \left( r,\varepsilon \right) }{f\left( \varepsilon \right) }%
\right) .
\end{equation}

Using the above equation and Eq. (\ref{Action2}), we have%
\begin{equation}
I=2\int \int_{0}^{H}\frac{dH}{\overset{.}{r_{h}}}dr_{h}=-2i\left[ \int
\int_{0}^{H}\frac{dH}{\left( \frac{g\left( \varepsilon \right) \psi \left(
r,\varepsilon \right) }{f\left( \varepsilon \right) }\right) }dr\right] .
\label{adi2}
\end{equation}

In order to solve this adiabatic invariant quantity (Eq. (\ref{adi2})), we
use the definition of Hawking's temperature, related to the surface gravity
on the outer horizon ($r_{+}$) as $T_{bh}=\frac{\hbar \kappa }{2\pi } $, in
which $\kappa $ is the surface gravity.

Inasmuch as the area spectrum and also the entropy spectrum spacing change
with respect to the change in coordinate transformation. In other words, the
adiabatic invariant quantity ($\int p_{i}dq_{i}$) used in Majhi and
Vagenas's method is not canonically invariant, hence Jiang and Han modified
this idea by considering the closed contour integral $\oint p_{i}dq_{i}$
which is invariant under coordinate transformations \cite{Jiang}. The closed
contour integral can be considered as a path that goes from $q_{i}^{out}$\
to $q_{i}^{in}$, in which $q_{i}^{out}$\ and $q_{i}^{in}$ are outside and
inside the event horizons, respectively. So, the adiabatic invariant
quantity is%
\begin{equation}
I=\oint
p_{i}dq_{i}=\int_{q_{i}^{in}}^{q_{i}^{out}}p_{i}^{out}dq_{i}+%
\int_{q_{i}^{out}}^{q_{i}^{in}}p_{i}^{in}dq_{i}.  \label{adi3}
\end{equation}%
where $p_{i}^{in}$ and $p_{i}^{out}$ are the conjugate momentums
corresponding to the coordinate $q_{i}^{in}$ and $q_{i}^{out}$,
respectively, and also $i=0,1,2,...$ . It is notable that, $%
q_{1}^{in}=r_{h}^{in}$, $q_{1}^{out}=r_{h}^{out}$ and also, $%
q_{0}^{in}=q_{0}^{out}=\tau $. Therefore, we can obtain the area spectrum of
this black hole by using the tunnelling method and the covariant action (\ref%
{adi3}). Entropy spectrum of various black holes have been studied in many
literatures \cite%
{EnspecM1,EnspecM2,EnspecM3,Enspec1,Enspec2,Enspec3,Enspec4,Enspec5,Enspec6,Enspec7,Enspec8,Enspec9,Enspec10}%
. In the following, we obtain entropy spectrum of BTZ black holes in massive
gravity's rainbow. Then, we investigate the effects such quantization on the
properties of the black holes.

\section{Entropy spectrum of BTZ black holes in massive gravity's rainbow}

The metric of $3$-dimensional spacetime in the presence of the gravity's
rainbow is given by
\begin{equation}
ds^{2}=-\frac{\psi (r,\varepsilon )}{f(\varepsilon )^{2}}dt^{2}+\frac{1}{%
g(\varepsilon )^{2}}\left( \frac{dr^{2}}{\psi (r,\varepsilon )}%
+r^{2}d\varphi ^{2}\right) ,  \label{metric}
\end{equation}%
in which $\psi (r,\varepsilon )$ is the metric function of our black holes
and $f(\varepsilon )$ and $g(\varepsilon )$ functions are rainbow functions.
The Lagrangian governing $3$-dimensional form of massive gravity is given by
\begin{equation*}
L_{massive}=m\left( \varepsilon \right) ^{2}\sum_{i=1}^{3}c_{i}(\varepsilon )%
\mathcal{U}_{i}(g,f),
\end{equation*}%
where $c(\varepsilon )_{i}$'s are some energy dependent constants and $U_{i}$%
's are symmetric polynomials of the eigenvalues of the $3\times 3$ matrix $%
K_{\nu }^{\mu }=\sqrt{g^{\mu \alpha }f_{\alpha \nu }}$ written as
\begin{equation*}
\mathcal{U}_{1}=\left[ \mathcal{K}\right] ,\;\;\;\;\;\mathcal{U}_{2}=\left[
\mathcal{K}\right] ^{2}-\left[ \mathcal{K}^{2}\right] ,\;\;\;\;\;\mathcal{U}%
_{3}=\left[ \mathcal{K}\right] ^{3}-3\left[ \mathcal{K}\right] \left[
\mathcal{K}^{2}\right] +2\left[ \mathcal{K}^{3}\right] ,
\end{equation*}%
which leads to following field equation
\begin{eqnarray}
\chi _{\mu \nu } &=&-\frac{c_{1}(\varepsilon )}{2}\left( \mathcal{U}%
_{1}g_{\mu \nu }-\mathcal{K}_{\mu \nu }\right) -\frac{c_{2}(\varepsilon )}{2}%
\left( \mathcal{U}_{2}g_{\mu \nu }-2\mathcal{U}_{1}\mathcal{K}_{\mu \nu }+2%
\mathcal{K}_{\mu \nu }^{2}\right)  \notag \\
&&-\frac{c_{3}(\varepsilon )}{2}(\mathcal{U}_{3}g_{\mu \nu }-3\mathcal{U}_{2}%
\mathcal{K}_{\mu \nu }+6\mathcal{U}_{1}\mathcal{K}_{\mu \nu }^{2}-6\mathcal{K%
}_{\mu \nu }^{3}).  \label{massiveTerm}
\end{eqnarray}

The only non-zero term of massive gravity is $\mathcal{U}_{1}$. Therefore,
the action for $3$-dimensional Einstein-massive-rainbow gravity in the
presence of Maxwell field is given by
\begin{equation}
\mathcal{I}=-\frac{1}{16\pi G(\varepsilon )}\int d^{3}x\sqrt{-g}\left[
\mathcal{R}-2\Lambda \left( \varepsilon \right) -\mathcal{F}+m\left(
\varepsilon \right) ^{2}c_{1}(\varepsilon )\mathcal{U}_{1}(g,f)\right] ,
\label{Action}
\end{equation}%
in which $R$ and $F$ are the scalar curvature and the Lagrangian of Maxwell
electrodynamics respectively. $G(\varepsilon )$ is gravitational constant
which is energy dependent. $\Lambda \left( \varepsilon \right) $ is the
energy dependent cosmological constant and $f$ and $g$ are a fixed symmetric
tensor and metric tensor, respectively. It is notable that, $m\left(
\varepsilon \right) $ is related to the energy dependent mass of graviton.
In addition, $F=F_{\mu \nu }F^{\mu \nu }$\ is the Maxwell invariant, in
which $F_{\mu \nu }=\partial _{\mu }A_{\nu }-\partial _{\nu }A_{\mu }$ is
the electromagnetic tensor with $A_{\mu }$ as its gauge potential. It is a
matter of calculation to show that field equations are obtained as
\begin{equation}
R_{\mu \nu }-\left( \frac{R}{2}-\Lambda \left( \varepsilon \right) \right)
g_{\mu \nu }+G\left( \varepsilon \right) \left( \frac{1}{2}g_{\mu \nu }%
\mathcal{F}-2L_{\mathcal{F}}F_{\mu \rho }F_{\nu }^{\rho }\right) +m\left(
\varepsilon \right) ^{2}\chi _{\mu \nu }=0,  \label{Field
    equation}
\end{equation}%
\begin{equation}
\partial _{\mu }\left( \sqrt{-g}F^{\mu \nu }\right) =0.
\label{Maxwell equation}
\end{equation}

The metric function is obtained in this gravity as \cite{BTZmassrain}%
\begin{equation}
\psi (r,\varepsilon )=-\frac{\Lambda \left( \varepsilon \right) r^{2}}{%
g(\varepsilon )^{2}}-m_{0}\left( \varepsilon \right) -2G\left( \varepsilon
\right) f(\varepsilon )^{2}q\left( \varepsilon \right) ^{2}\ln \left( \frac{r%
}{l(\varepsilon )}\right) +\frac{m\left( \varepsilon \right)
^{2}c(\varepsilon )c_{1}(\varepsilon )r}{g(\varepsilon )^{2}},
\label{f(r)ENMax2}
\end{equation}%
where $m_{0}(\varepsilon )$ is an energy dependent integration constant
related to the total mass of the BTZ black holes.

The electric potential ($U$) and the total electric charge ($Q$) are
calculated as \cite{BTZmassrain}%
\begin{eqnarray}
U\left( \varepsilon \right) &=&-q\left( \varepsilon \right) \ln \left( \frac{%
r_{+}}{l(\varepsilon )}\right) .  \label{TotalU} \\
&&  \notag \\
Q\left( \varepsilon \right) &=&\frac{1}{2}f\left( \varepsilon \right)
G\left( \varepsilon \right) q\left( \varepsilon \right) .  \label{TotalQ}
\end{eqnarray}

Using the standard definition of the Hawking temperature ($T=\frac{\hbar
\kappa }{2\pi }$), the surface gravity is obtained by considering the metric
(\ref{metric}) as \cite{BTZmassrain}%
\begin{equation}
\kappa =\frac{1}{2\pi }\sqrt{\frac{-1}{2}\left( \nabla _{\mu }\chi _{\nu
}\right) \left( \nabla ^{\mu }\chi ^{\nu }\right) }=\frac{1}{2}\left( \frac{%
g\left( \varepsilon \right) \psi ^{\prime }(r,\varepsilon )}{f\left(
\varepsilon \right) }\right) .  \label{kappa}
\end{equation}

Therefore, the Hawking's temperature of these black holes are \cite%
{BTZmassrain}%
\begin{equation}
T=\frac{\hbar \kappa }{2\pi }=\frac{\hbar }{4\pi }\left( \frac{g\left(
\varepsilon \right) \psi ^{\prime }(r,\varepsilon )}{f\left( \varepsilon
\right) }\right) \left\vert _{r=r_{+}}\right. =-\frac{\Lambda \left(
\varepsilon \right) r_{+}}{2\pi f\left( \varepsilon \right) g\left(
\varepsilon \right) }+\frac{m\left( \varepsilon \right) ^{2}c(\varepsilon
)c_{1}(\varepsilon )}{4\pi f\left( \varepsilon \right) g\left( \varepsilon
\right) }-\frac{f\left( \varepsilon \right) g\left( \varepsilon \right)
G\left( \varepsilon \right) q\left( \varepsilon \right) ^{2}}{2\pi r_{+}} ,
\label{temp}
\end{equation}%
where $r_{+}$ is outer horizon of black hole. The entropy of black holes can
be obtained by employing the area law as \cite{BTZmassrain}
\begin{equation}
S=\frac{\pi r_{+}}{2g\left( \varepsilon \right) }.  \label{TotalS}
\end{equation}

The total mass of these solutions is given by \cite{BTZmassrain}
\begin{equation}
M=\frac{m_{0}\left( \varepsilon \right) }{8f\left( \varepsilon \right) }.
\label{TotalM}
\end{equation}

Here, we want to quantize the entropy of this black hole using the adiabatic
invariant quantity and Bohr-Sommerfeld quantization rule. Considering Eqs. (%
\ref{adi2}) and (\ref{adi3}), we have%
\begin{equation}
I=\oint p_{i}dq_{i}=-4i\left[ \int_{r_{out}}^{r_{in}}\int_{0}^{H}\frac{dH}{%
\psi \left( r,\varepsilon \right) }dr\right] \times \frac{f\left(
\varepsilon \right) }{g\left( \varepsilon \right) }.  \label{adi4}
\end{equation}

In order to solve the above equation, we use the near horizon approximation,
so $\psi \left( r\right) $ can be Taylor expanded in the following form%
\begin{equation}
\psi \left( r,\varepsilon \right) =\psi \left( r,\varepsilon \right)
_{r=r_{+}}+\left( r-r_{+}\right) \psi ^{\prime }(r,\varepsilon
)_{r=r_{+}}+...~.
\end{equation}

The first term is zero ($\psi \left( r,\varepsilon \right) _{r=r_{+}}=0$).
Using the Cauchy integral theorem and temperature (\ref{temp}), Eq. (\ref%
{adi4}) reduces to%
\begin{equation}
I=\oint p_{i}dq_{i}=4\pi \int_{0}^{H}\frac{dH}{\kappa }=2\hbar \int_{0}^{H}%
\frac{dH}{T}.  \label{adi5}
\end{equation}

The Smarr-formula for BTZ black hole in massive gravity's rainbow is
\begin{equation}
dM=dH=TdS-UdQ.
\end{equation}

Therefore, the equation (\ref{adi5}) become%
\begin{equation}
\oint p_{i}dq_{i}=2\hbar S\left[ 1+\frac{U(\varepsilon )f\left( \varepsilon
\right) G\left( \varepsilon \right) }{2Q\left( \varepsilon \right) }\ln
\left( G\left( \varepsilon \right) \left[ 2\Lambda \left( \varepsilon
\right) r_{+}-m^{2}\left( \varepsilon \right) c\left( \varepsilon \right)
c_{1}\left( \varepsilon \right) \right] r_{+}+8Q^{2}\left( \varepsilon
\right) g^{2}\left( \varepsilon \right) \right) \right] .  \label{adi6}
\end{equation}

On the other hand, the Bohr-Sommerfeld quantization rule is given by%
\begin{equation}
\oint p_{i}dq_{i}=2\pi n\hbar ,\ \ \ \ \ \ \ n=1,2,3,...\ .  \label{adi7}
\end{equation}

Comparing Eq. (\ref{adi6}) with Eq. (\ref{adi7}), one can obtain the entropy
spectrum as%
\begin{equation}
S=\frac{n\pi }{1+\frac{U(\varepsilon )f\left( \varepsilon \right) G\left(
\varepsilon \right) }{2Q\left( \varepsilon \right) }\ln \{G\left(
\varepsilon \right) \left[ 2\Lambda \left( \varepsilon \right)
r_{+}-m^{2}\left( \varepsilon \right) c\left( \varepsilon \right)
c_{1}\left( \varepsilon \right) \right] r_{+}+8Q^{2}\left( \varepsilon
\right) g^{2}\left( \varepsilon \right) \}}.
\end{equation}

Quantization of the entropy has specific physical results which are:

I) the quantization results into formation of a spectrum of the entropy
characterized by $n$. The entropy is an increasing function of the $n$, but
the general behavior of the entropy is not determined by this parameter.

II) the entropy spectrum is a function of black hole's properties (the
electric field, the massive parameters, the cosmological constant and
gravity's rainbow generalizations and horizon radius).

\begin{figure}[!htb]
\centering
\includegraphics[width=0.3\linewidth]{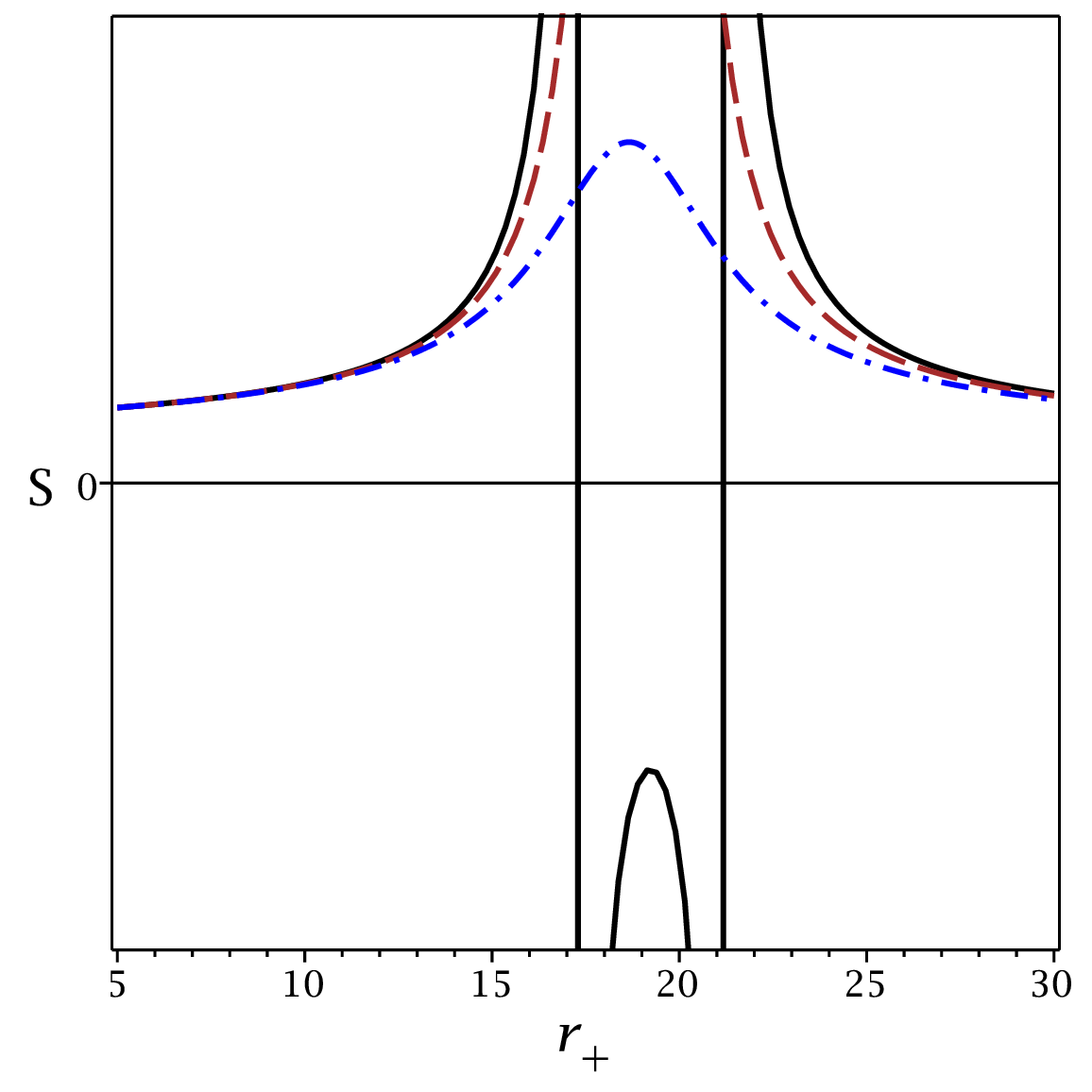}
\par
\medskip \includegraphics[width=0.3\linewidth]{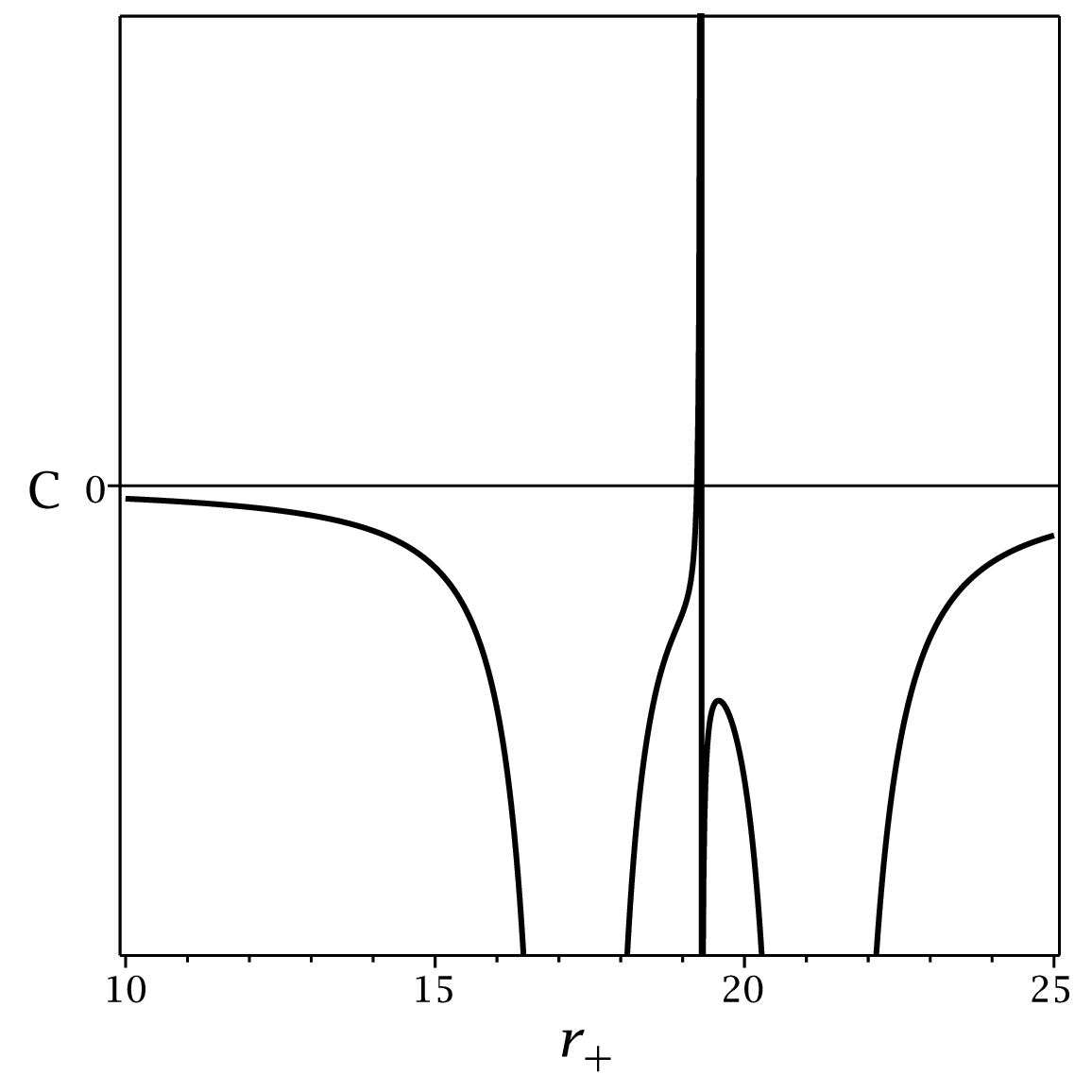}\hfil
\includegraphics[width=0.3\linewidth]{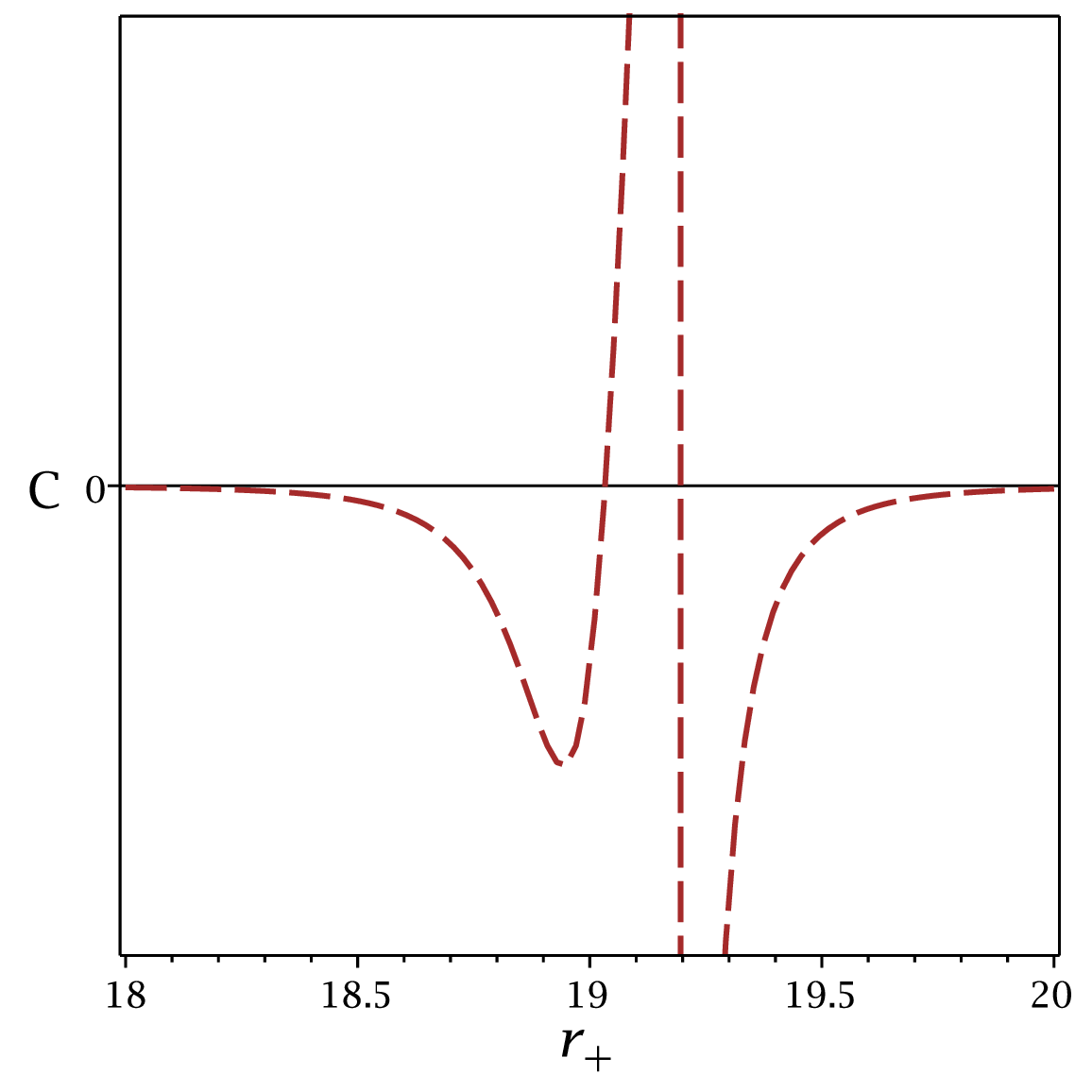}\hfil
\includegraphics[width=0.3\linewidth]{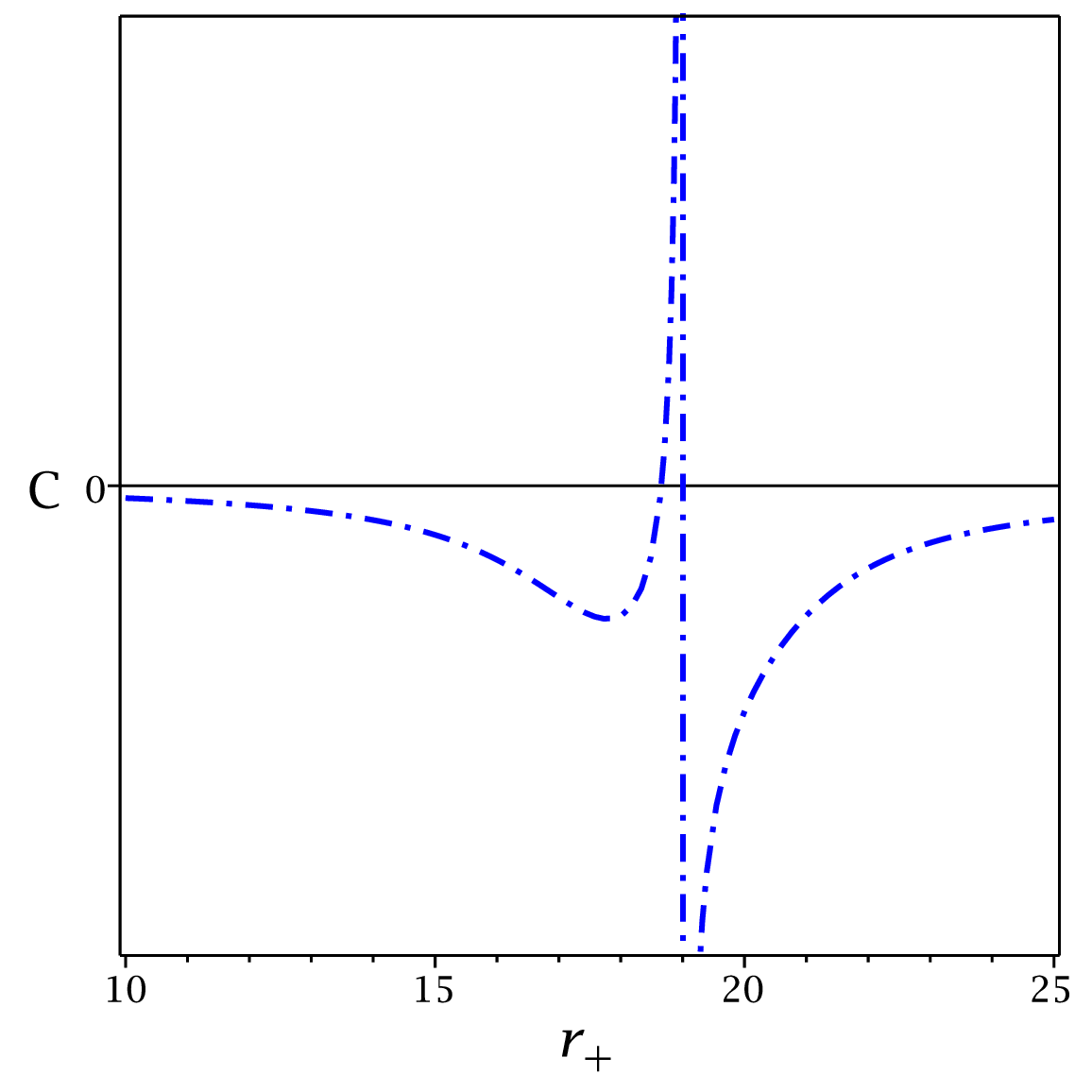}
\caption{$S$ (upper panel) and $C$ (lower panels) versus $r_{+}$ for $G(%
\protect\varepsilon )=Q (\protect\varepsilon)=U (\protect\varepsilon)=c(%
\protect\varepsilon)=c_{1}(\protect\varepsilon)=m(\protect\varepsilon)=n=1$,
$g(\protect\varepsilon)=f(\protect\varepsilon)=1.1$, $\Lambda(\protect%
\varepsilon)=0.0130$ (continuous line), $\Lambda(\protect\varepsilon)=0.0131
$ (dashed line) and $\Lambda(\protect\varepsilon)=0.0134$ (dashed-dotted
line).}
\label{Fig1}
\end{figure}

III) while the usual entropy of black holes (\ref{TotalS}) is divergent free
and smooth function of the horizon radius, the quantization results into the
possibility of divergencies for the entropy. The divergent points of entropy
are obtained as
\begin{equation}
\left. r_{+}\right\vert _{S\rightarrow \infty }=\frac{G(\varepsilon
)c(\varepsilon )c_{1}(\varepsilon )m^{2}(\varepsilon )\pm \sqrt{%
G^{2}(\varepsilon )c^{2}(\varepsilon )c_{1}^{2}(\varepsilon
)m^{4}(\varepsilon )-64G(\varepsilon )\Lambda (\varepsilon
)Q^{2}(\varepsilon )g^{2}(\varepsilon )+8G(\varepsilon )\Lambda (\varepsilon
)e^{-\frac{2Q\left( \varepsilon \right) }{U(\varepsilon )f\left( \varepsilon
\right) G\left( \varepsilon \right) }}}}{4G(\varepsilon )\Lambda
(\varepsilon )},  \label{div}
\end{equation}%
which shows that under certain conditions, the quantized entropy could have
up to two divergencies (Fig. \ref{Fig1}). It should be noted that only
positive values of ($\ref{div}$) are physically acceptable. In the absence
of massive gravity, only for AdS solutions, divergent entropy could be
obtained. In general, the major condition for existence of the divergent
quantized entropy is given by
\begin{equation}
\Lambda (\varepsilon )\leq \frac{G(\varepsilon )c^{2}(\varepsilon
)c_{1}^{2}(\varepsilon )m^{4}(\varepsilon )}{64Q^{2}(\varepsilon
)g^{2}(\varepsilon )-e^{-\frac{2Q\left( \varepsilon \right) }{U(\varepsilon
)f\left( \varepsilon \right) G\left( \varepsilon \right) }}},
\end{equation}%
with positivity of ($\ref{div}$). If one consider absence of the divergency
in entropy as a requirement for having physical solutions, the mentioned
condition gives us an upper (lower) limit on massive gravity's parameter
(the cosmological constant).

IV) quantization of the entropy is also could provide the possibility of
formation of root for this quantity. The root of entropy is given by
\begin{equation}
\left. r_{+}\right\vert _{S=0}=\frac{G(\varepsilon )c(\varepsilon
)c_{1}(\varepsilon )m^{2}(\varepsilon )\pm \sqrt{G^{2}(\varepsilon
)c^{2}(\varepsilon )c_{1}^{2}(\varepsilon )m^{4}(\varepsilon
)-64G(\varepsilon )\Lambda (\varepsilon )Q^{2}(\varepsilon
)g^{2}(\varepsilon )}}{4G(\varepsilon )\Lambda (\varepsilon )},  \label{root}
\end{equation}%
which shows that the existence of root for entropy is restricted to
satisfaction of the following condition
\begin{equation}
\Lambda (\varepsilon )\leq \frac{G(\varepsilon )c^{2}(\varepsilon
)c_{1}^{2}(\varepsilon )m^{4}(\varepsilon )}{64Q^{2}(\varepsilon
)g^{2}(\varepsilon )},
\end{equation}%
and positivity of (\ref{root}). It should be noted that for usual entropy (%
\ref{TotalS}), the only root for the entropy exists at $r_{+}=0$, whereas
for the quantized entropy, the root is modified to a non-zero horizon radius.

V) if the following relation holds
\begin{equation}
\left. r_{+}\right\vert _{\text{free}}=\frac{G(\varepsilon )c(\varepsilon
)c_{1}(\varepsilon )m^{2}(\varepsilon )\pm \sqrt{G^{2}(\varepsilon
)c^{2}(\varepsilon )c_{1}^{2}(\varepsilon )m^{4}(\varepsilon
)-64G(\varepsilon )\Lambda (\varepsilon )Q^{2}(\varepsilon
)g^{2}(\varepsilon )+8G(\varepsilon )\Lambda (\varepsilon )}}{4G(\varepsilon
)\Lambda (\varepsilon )},  \label{free}
\end{equation}%
the quantized entropy will be independent of black hole's properties. In
other words, the quantized entropy will have a fixed value irrespective of
variations in black hole's electric charge, the massive parameters, the
cosmological constant and horizon radius. This is one of the consequences of
quantization of entropy which says that the entropy will be independent of
the size and electric charge of the black holes (if Eq. (\ref{free}) holds).

VI) one of the most important results of the quantization is non-zero
entropy for $r_{+}=0$ and is given by
\begin{equation}
S=\frac{n\pi }{1+\frac{U(\varepsilon )f\left( \varepsilon \right) G\left(
\varepsilon \right) }{2Q\left( \varepsilon \right) }\ln \{8Q^{2}\left(
\varepsilon \right) g^{2}\left( \varepsilon \right) \}}.
\end{equation}

The limit $r_{+}\rightarrow 0$ is known as high energy limit. Evidently, in
this limit, the quantized entropy is non-zero (in contrast to usual entropy)
and it is governed by the electric part of black holes, gravity's rainbow
generalization and $n$. Another interpretation of this limit is that for
evaporation of the black holes, despite the vanishing internal energy of the
black holes, the entropy remains non-zero. Interestingly, in this limit, a
non-zero temperature could be also observed, but while the entropy in this
limit is independent of massive gravity, the temperature only depends on
massive gravity \cite{BTZmassrain}.

The asymptotic behavior of quantized entropy is given by
\begin{equation}
\lim_{r_{+}\rightarrow \infty }S=\frac{n\pi }{1+\frac{U(\varepsilon )f\left(
\varepsilon \right) G\left( \varepsilon \right) }{2Q\left( \varepsilon
\right) }\ln \{2G\left( \varepsilon \right) \Lambda \left( \varepsilon
\right) r_{+}^{2}\}}+O(\frac{1}{r_{+}}),
\end{equation}%
which shows that in this limit, the only non-contributing factor on the
behavior of entropy is the massive gravity.

To further clarifies the effects of quantization on thermodynamic of the
black holes, we investigate the heat capacity. The heat capacity gives a
detailed picture regarding thermal/thermodynamical behavior of the
solutions. In general, for this black holes with quantized entropy, this
quantity is given by
\begin{equation}
C=T\frac{\left( \frac{\partial S}{\partial r_{+}}\right) _{Q,U}}{\left(
\frac{\partial T}{\partial r_{+}}\right) _{Q,U}}={\frac{n\pi \,Q\left(
\varepsilon \right) U(\varepsilon )f\left( \varepsilon \right) G\left(
\varepsilon \right) ^{2}\left( m^{2}\left( \varepsilon \right) c\left(
\varepsilon \right) c_{1}\left( \varepsilon \right) -4\,\Lambda \left(
\varepsilon \right) r_{+}\right) r_{+}}{\left[ 2\,Q\left( \varepsilon
\right) +UfG\ln \{G\left( \varepsilon \right) \left[ 2\Lambda \left(
\varepsilon \right) r_{+}-m^{2}\left( \varepsilon \right) c\left(
\varepsilon \right) c_{1}\left( \varepsilon \right) \right]
r_{+}+8Q^{2}\left( \varepsilon \right) g^{2}\left( \varepsilon \right) \}%
\right] ^{2}Z},}  \label{heat}
\end{equation}%
where $Z=G\left( \varepsilon \right) {r}_{+}^{2}\Lambda \left( \varepsilon
\right) -4\,Q\left( \varepsilon \right) ^{2}g^{2}\left( \varepsilon \right) $%
.

Evidently, by quantizing the entropy, the heat capacity is consequently
quantized. But here, we should be a little bit cautious. The reason is that
quantization is only done for the entropy while the temperature is not
quantized. In addition, we have considered an ensemble where the electric
charge and the potential are both fixed (canonical ensemble). The obtained
heat capacity highlights several important contributions of quantized
entropy:

I) here as well, due to quantization, a spectrum is formed by the heat
capacity characterized by $n$. But overall, the behavior of heat capacity is
not determined by $n$.

II) the positivity/negativity of heat capacity determines thermal
stability/instability of the solutions. Therefore, the stability conditions
are given by following set of conditions

\begin{equation}
\left\{
\begin{array}{c}
\frac{4\,Q\left( \varepsilon \right) ^{2}g^{2}\left( \varepsilon \right) }{%
G\left( \varepsilon \right) {r}_{+}^{2}}<\Lambda \left( \varepsilon \right) <%
\frac{m^{2}\left( \varepsilon \right) c\left( \varepsilon \right)
c_{1}\left( \varepsilon \right) }{4\,r_{+}} \\
\\
\frac{4\,Q\left( \varepsilon \right) ^{2}g^{2}\left( \varepsilon \right) }{%
G\left( \varepsilon \right) {r}_{+}^{2}}>\Lambda \left( \varepsilon \right) >%
\frac{m^{2}\left( \varepsilon \right) c\left( \varepsilon \right)
c_{1}\left( \varepsilon \right) }{4\,r_{+}}%
\end{array}%
\right. ,
\end{equation}%
which confirms that for asymptotical flat and AdS solutions, the heat
capacity will be negative and solutions will always be thermally unstable.
The only possible thermally stable solution exists for dS branch and under
satisfaction of certain conditions.

III) the root of heat capacity is given by

\begin{equation}
\left. r_{+}\right\vert _{C=0}=\left\{
\begin{array}{c}
\frac{m^{2}\left( \varepsilon \right) c\left( \varepsilon \right)
c_{1}\left( \varepsilon \right) }{4\,\Lambda \left( \varepsilon \right) } \\
\\
\frac{G(\varepsilon )c(\varepsilon )c_{1}(\varepsilon )m^{2}(\varepsilon
)\pm \sqrt{G^{2}(\varepsilon )c^{2}(\varepsilon )c_{1}^{2}(\varepsilon
)m^{4}(\varepsilon )-64G(\varepsilon )\Lambda (\varepsilon
)Q^{2}(\varepsilon )g^{2}(\varepsilon )}}{4G(\varepsilon )\Lambda
(\varepsilon )}%
\end{array}%
\right. ,  \label{rootC}
\end{equation}%
which confirms two important points: first of all, one of the possible roots
is originated only from contribution of the massive gravity. The second
point is that some of the roots of heat capacity are also entropy's roots
(please compare Eqs. (\ref{root}) and (\ref{rootC})).

IV) the divergencies in heat capacity are where thermal phase transitions
take place. The divergencies of heat capacity are given by

\begin{equation}
\left. r_{+}\right\vert _{C\rightarrow \infty }=\left\{
\begin{array}{c}
\frac{2Q\left( \varepsilon \right) g\left( \varepsilon \right) }{\sqrt{%
\,G\left( \varepsilon \right) \Lambda \left( \varepsilon \right) }} \\
\\
\frac{G(\varepsilon )c(\varepsilon )c_{1}(\varepsilon )m^{2}(\varepsilon
)\pm \sqrt{G^{2}(\varepsilon )c^{2}(\varepsilon )c_{1}^{2}(\varepsilon
)m^{4}(\varepsilon )-64G(\varepsilon )\Lambda (\varepsilon
)Q^{2}(\varepsilon )g^{2}(\varepsilon )+8G(\varepsilon )\Lambda (\varepsilon
)e^{-\frac{2Q\left( \varepsilon \right) }{U(\varepsilon )f\left( \varepsilon
\right) G\left( \varepsilon \right) }}}}{4G(\varepsilon )\Lambda
(\varepsilon )}%
\end{array}%
\right. .  \label{divC}
\end{equation}

The quantization has also resulted into modification in place and number of
the divergencies in heat capacity comparing to non-quantized case.
Evidently, it is possible for the heat capacity to have up to three
divergencies (see right panel of Fig. \ref{Fig1}). One of these divergencies
is due to contributions and interactions of the cosmological constant and
the electric charge. The other divergencies are same as the divergencies
obtained for the entropy (please compare Eqs. (\ref{div}) and (\ref{divC}))
indicating that these phase transitions are due to the quantization. In
other words, through the quantization of entropy, novel thermal phase
transitions are introduced in thermodynamical structure of the black holes.

\section{Conclusions}

In this paper, we considered the BTZ black holes in the presence of massive
gravity's rainbow. We studied the quantization of entropy of these black
hole using an adiabatic invariant integral method put forwarded by Majhi and
Vagenas with modification proposed by Jiang and Han, and the Bohr-Sommerfeld
quantization rule.

It was shown that quantization of the entropy results into formation of a
spectrum of entropy. In addition, the quantized entropy leads to the
existence of divergencies and roots for the entropy which were absent in the
usual entropy. It was also shown that in the high energy limit and/or in
vanishing horizon radius, the entropy has a non-zero value which again was
in contrast to the usual entropy. In general, the behavior of quantized
entropy depends on the parameter of horizon radius, massive gravity, rainbow
functions, the cosmological and the Newton constants, and also the electric
charge. But it was shown that for specific choices of different parameters,
the effects of the black holes' properties (both gravitational and matter
field contributions) could be cancelled resulting into a fixed entropy. In
other words, for this specific case, the quantized entropy of black holes
was independent of the size, electric field and other characteristics of the
black holes. In addition, the heat capacity of solutions was investigated.
It was shown that quantization resulted into appearance of novel phase
transition points into structure of the black holes. Also, it was shown that
due to quantization, only the dS black holes could be thermally stable while
asymptotically flat and AdS solutions were unstable.

\begin{acknowledgements} BE and SHH thank Shiraz University Research Council. This work has been supported financially by the Research Institute for
Astronomy and Astrophysics of Maragha, Iran. \end{acknowledgements}

\end{document}